\documentstyle[12pt]{article}
\begin{document}
\begin{titlepage}
\mbox{ }
\rightline{UCT-TP-207/94}
\rightline{April 1994}
\vspace{1.5cm}
\begin{center}
\begin{Large}
{\bf  Mass of the charm-quark from QCD sum rules}

\end{Large}

\vspace{2cm}

{\large {\bf C. A. Dominguez \footnote{John Simon Guggenheim Fellow 1994-1995}
and G. R. Gluckman}}

Institute of Theoretical Physics and Astrophysics, University of
Cape Town, Rondebosch 7700, South Africa\\

\vspace{3mm}

and \\

\vspace{3mm}

{\large {\bf N. Paver}}\\[0.7cm]
Dipartimento di Fisica Teorica, Universit\`{a} di Trieste, Italy \\
\vspace{3mm}
Istituto Nazionale di Fisica Nucleare, Sezione di Trieste, Italy\\
\end{center}

\vspace{1.2cm}

\begin{abstract}
\noindent
Relativistic and non-relativistic ratios of Laplace transform QCD moment
sum rules  for charmonium
are used in order to determine the value of the on-shell charm-quark
mass. The validity of the
non-relativistic version of QCD sum rules in this particular application is
discussed. After using current values of the perturbative and
non-perturbative QCD parameters, as well as experimental data on the
$J/\psi$ system, we obtain $m_{c}(Q^{2}=m_{c}^{2}) = 1.46 \pm 0.07
\;\mbox{GeV}$.
\end{abstract}
\end{titlepage}

\setlength{\baselineskip}{1.5\baselineskip}
\noindent

Recently, the value of the (on-shell) beauty-quark mass has been
determined \cite{ref1} by confronting very accurate experimental data
on the upsilon system \cite {ref2} with ratios of  non-relativistic
Laplace transform QCD moments.
This theoretical framework, suggested by Bertlmann \cite {ref3},
offers several advantages, e.g. radiative and non-perturbative
corrections are well under control, and the non-relativistic limit
follows quite naturally  from quantum mechanical analogues \cite {ref4}.
This version of QCD sum rules leads to an expansion in powers of the inverse
of the heavy quark mass which allows one to test the range of validity of the
non-relativistic limit, and more generally, to assess the role of
mass corrections. This might be of interest for calculations based on the
simplifying assumption $\Lambda_{QCD}/m_Q\ll 1$.

Non-relativistic Laplace moments appear to have a sensitive dependence on the
quark mass. In fact, in spite of the large uncertainties
affecting the values of $\Lambda_{QCD}$ and the non-perturbative gluon
condensate, $m_{b}$ can be extracted from the upsilon data with high
precision. This extraction is performed by confronting the ratios of
Laplace transform moments calculated from experiment with those from theory.
The latter involve the QCD parameters $m_{b}$, $\Lambda$, $<\alpha_{s}
G^{2}>$, etc.. These ratios are functions of the Laplace variable, which
acts as a short distance expansion parameter, and one finds a reasonably
wide region in this variable where there is a matching between experiment
and theory for a specific value of the quark mass.

As pointed out in \cite{ref1}, a straightforward extension of
this technique to the charm-quark may not work, as radiative and mass
corrections could exceed 100 \% . This would be true if the window in the
Laplace variable would be the same for beauty and for charm. However, there
is no {\it a-priori} reason for this to be the case. In fact, as also
suggested in \cite{ref3}, the matching between theory and experiment for
the beauty and the charm quarks could take place at different ranges of the
Laplace variable. If this range is such that radiative, non-perturbative, and
mass corrections remain small, then it would become possible to extract the
value of the charm quark from this framework.

In this note we  study the ratios of relativistic Laplace transform
QCD moments for the case of charm, and compare them with the
non-relativistic versions. This provides a measure of the validity of the
heavy quark mass expansion in this particular application, as well as an
estimate of the systematic uncertainties affecting this technique. We make
use of the current values of $\Lambda$ and  $<\alpha_{s} G^{2}>$, and carry
out the non-relativistic expansion of the Laplace ratios to
next-to-next to leading order
in $1/m_{c}$. We find this expansion to converge reasonably fast, and
the radiative and non-perturbative contributions  to be safely under control.
Thanks to this feature and to the high accuracy of the experimental data,
as in the case of $m_b$, the extracted value of the (on-shell)
charm-quark mass is affected by a relatively small uncertainty, in spite of
the large uncertainties in $\Lambda$ and  $<\alpha_{s} G^{2}>$. In
connection with the latter, we recall that most of the early (relativistic)
QCD sum rule analyses of charmonium attempted to extract the values of both the
charm-quark mass and the gluon condensate. Since $<\alpha_{s} G^{2}>$ is now
known independently from $e^{+} e^{-}$ and from $\tau$ decay data \cite{ref5},
one is in a better position to
determine $m_{c}$. In fact, for a given pair of values of
$\Lambda$ and  $<\alpha_{s} G^{2}>$, the matching between theory and
experiment becomes a one parameter fit, the parameter being $m_{c}$.
We find that the values of $m_{c}$ from the
fully relativistic ratios are in good agreement (within errors) with those
from the non-relativistic ratios.
We conclude with a  comparison
of our results with those obtained previously by other authors.

We begin by considering the two-point function
\begin{equation}
\Pi_{\mu \nu} (q) =  i \; \int d^{4}x \exp (i q x) \langle0|T(V_{\mu} (x)
V^{+}_{\nu}
(0))|0 \rangle
		   =  (- g_{\mu \nu} q^{2} + q_{\mu} q_{\nu}) \Pi
		  (q^{2}) \; ,
\end{equation}
with $V_{\mu}(x)=\bar{c}(x) \gamma_{\mu} c(x)$. The function $\Pi(q^{2})$
has been calculated in perturbative QCD at the two-loop level \cite{ref6},
with its imaginary part given by
\begin{equation}
\frac{1}{\pi} \; \mbox{Im} \; \Pi (s)|_{QCD} = \frac{1}{8 \pi^{2}}
v(3 - v^{2})
\Biggl\{ 1 + \frac{4 \alpha_{s}}{3} \left[ \frac{\pi}{2v} - \frac{(v+3)}{4}
\left( \frac{\pi}{2} - \frac{3}{4 \pi} \right) \right] \Biggr\}
\theta (s - 4 m_{c}^{2}) \; ,
\end{equation}

where $v=\sqrt{1 - 4 m_{c}^{2}/s}$, and $m_{c}$ is the charm-quark
on-shell mass:  $m_{c}=m_{c}(Q^{2}=m_{c}^{2})$. The leading
non-perturbative term in the operator product expansion of   $\Pi(q^{2})$
involves the gluon condensate, i.e.

\begin{equation}
\Pi (s)|_{NP} = \frac{1}{48 \; s^{2}} \times \left[
\frac{3(v^{2} + 1)(1 - v^{2})^{2}}{2 v^{5}} \ln \frac{1+v}{1-v}
- \frac{3v^{4} - 2v^{2} + 3}{v^{4}} \right]
\langle \frac{\alpha_{s}}{\pi} G^{2} \rangle \; .
\end{equation}

The function $\Pi(q^{2})$ satisfies a once-subtracted dispersion relation,
and the subtraction constant can be disposed of by taking the Laplace
transform
\begin{equation}
\Pi (\sigma) =  \int_{0}^{\infty} ds \exp (- \sigma s)
\; \mbox{Im} \; \Pi (s)   \; .
\end{equation}
The quantity of interest to us here is the ratio of the first two Laplace
moments, which can be expressed as
\begin{equation}
{\cal{R}} (\sigma) = - \frac{d}{d \sigma} \ln \; \Pi (\sigma) \; .
\end{equation}

{}From (2) and (3) one  obtains \cite{ref3}
\begin{equation}
\Pi (\sigma) = \exp (- 4 m_{c}^{2} \sigma) \pi \; A(\sigma) [1 +
a(\sigma) \; \alpha_{s} + b(\sigma) \phi] \; ,
\end{equation}
where, with $\omega = 4 m_{c}^{2} \sigma$,
\begin{equation}
\pi A(\omega) = \frac{3}{16 \sqrt{\pi}} \frac {4 m_{c}^{2}}{\omega}
\; G(\frac{1}{2}, \frac{5}{2}, \omega) \; ,
\end{equation}

\begin{equation}
a(\omega) = \frac{4}{3\sqrt{\pi}} \; G^{-1} (\frac{1}{2}, \frac{5}{2},
\omega)
[\pi - c_{1} \; G(1,2, \omega) + \frac{1}{3} c_{2} \; G(2,3,\omega)]
- c_{2} \; ,
\end{equation}

\begin{equation}
b(\omega) = - \frac{\omega^{2}}{2} \; G(-\frac{1}{2}, \frac{3}{2}, \omega)
\; G^{-1}(\frac{1}{2}, \frac{5}{2}, \omega) \; ,
\end{equation}

\begin{equation}
c_{1} = \frac{\pi}{3} + \frac{c_{2}}{2} \; \; \; , \; \; \;
c_{2} = \frac{\pi}{2} - \frac{3}{4 \pi} \; ,
\end{equation}

\begin{equation}
\alpha_{s} (Q^{2}) = \frac{12 \pi}{25 \; \ln \; Q^{2}/\Lambda^{2}} \; ,
\end{equation}

\begin{equation}
\phi = \frac{\pi}{36} \frac{<\alpha_{s} G^{2}>}{m_{c}^{4}} \; ,
\end{equation}

\begin{equation}
G(b,c,\omega) = \frac{\omega^{-b}}{\Gamma (c)} \int_{0}^{\infty}
dt \; t^{c-1} e^{-t} (1 + \frac{t}{\omega})^{-b} \; .
\end{equation}

The function $G(b,c,\omega)$ is related to the Whittaker function
$W_{\lambda,\mu}(\omega)$ through \cite{ref7}
\begin{equation}
G(b,c,\omega) = \omega^{\mu - 1/2} e^{\omega/2} \; W_{\lambda,\mu}
(\omega) \; ,
\end{equation}
with $\mu=(c-b)/2$, and $\lambda =(1-c-b)/2$. The ratio (5) can then be
calculated, with the result \cite{ref3}
\begin{equation}
{\cal{R}} (\omega) = 4 m_{c}^{2} \left[ 1 - \frac{A'(\omega)}{A(\omega)} -
\frac{a'(\omega) \alpha_{s} + b'(\omega) \phi}
{1 + a(\omega) \alpha_{s} + b (\omega) \phi} \right] \; ,
\end{equation}
where

\begin{equation}
A'(\omega) = - \frac{A(\omega)}{\omega} \left[ \frac{3}{2} -
\frac{5}{4} G(\frac{3}{2}, \frac{7}{2}, \omega) \; G^{-1} (\frac{1}{2},
\frac{5}{2}, \omega) \right] \; ,
\end{equation}

\begin{eqnarray*}
a'(\omega) = \frac{4}{3 \omega \sqrt{\pi}} \; G^{-1} (\frac{1}{2},
\frac{5}{2}, \omega) \Biggl\{ \frac{1}{2} \; G^{-1} (\frac{1}{2},
\frac{5}{2},
\omega) \; \left[ G(\frac{1}{2}, \frac{5}{2}, \omega)  \right. \Biggr.
\end{eqnarray*}
\begin{eqnarray*}
\left. - \frac{5}{2} G( \frac{3}{2}, \frac{7}{2}, \omega)
\right] \left[ \pi - c_{1} G(1,2, \omega) + \frac{c_{2}}{3}
G(2,3, \omega) \right]
\end{eqnarray*}
\begin{equation}
\left. + c_{1} [G(1,2, \omega) - 2 G(2,3, \omega)] + \frac{1}{3} c_{2}
[- 2 G(2,3, \omega) + 6 G(3,4, \omega)] \right\} \; ,
\end{equation}
\newpage
\begin{eqnarray*}
b'(\omega) = \frac{2}{\omega} b(\omega) - \frac{\omega}{4}
\left\{ G(- \frac{1}{2}, \frac{3}{2}, \omega)
G^{-1} (\frac{1}{2}, \frac{5}{2}, \omega) \right.
\end{eqnarray*}
\begin{equation}
\left.  - \frac{3}{2} + G(- \frac{1}{2}, \frac{3}{2}, \omega)
G^{-2} (\frac{1}{2}, \frac{5}{2}, \omega)
\left[ G(\frac{1}{2}, \frac{5}{2}, \omega) - \frac{5}{2}
G(\frac{3}{2}, \frac{7}{2}, \omega)  \right] \right\} \; .
\end{equation}

The above expressions (15)-(18) involve no approximations, other than the
two-loop perturbative expansion, and retaining the leading non-perturbative
term in the operator product expansion. We shall refer to Eq.(15) as the
fully relativistic Laplace ratio.

In the non-relativistic (heavy quark-mass) limit, the Laplace transform (4)
becomes
\begin{equation}
\Pi(\tau) =  \int_{0}^{\infty} dE \exp (- \tau E)
\; \mbox{Im} \;
\Pi(E) \; ,
\end{equation}

where $\tau = 4 m_{c} \sigma$, and $s = (2 m_{c} + E)^{2}$ so that
$E \geq 0$. The Laplace ratio (5) is now given by
\begin{equation}
{\cal{R}}(\tau) = 2 m_{c} - \frac{d}{d \tau} \; \ln \; \Pi(\tau) \; .
\end{equation}
After expanding the functions $G(b,c,\omega)$ entering Eqs.(7)-(9), and
(16)-(18), we obtain the non-relativistic ratio
\begin{eqnarray*}
{\cal{R}}(\tau) = 2 m_{c} \left\{ 1 + \frac{3}{4} \frac{1}{m_{c} \tau}
\left( 1 - \frac{5}{6} \frac{1}{m_{c} \tau} + \frac{10}{3}
\frac{1}{m_{c}^{2} \tau^{2}} \right) \right.
\end{eqnarray*}
\begin{eqnarray*}
- \frac{\sqrt{\pi}}{3} \frac{\alpha_{s}}{\sqrt{m_{c} \tau}}
\left[ 1 - \left( \frac{2}{3} + \frac{3}{8 \pi^{2}} \right)
\frac{1}{m_{c} \tau} + \frac{1}{32} \left(107 + \frac{51}{\pi^{2}}
\right) \frac{1}{m_{c}^{2} \tau^{2}} \right]
\end{eqnarray*}
\begin{equation}
\left.
+ \frac{\pi}{48} \frac{\tau^{2}}{m_{c}^{2}} \langle \alpha_{s} G^{2}
\rangle \left( 1 + \frac{4}{3} \frac{1}{m_{c} \tau} - \frac{5}{12}
\frac{1}{m_{c}^{2} \tau^{2}} \right) \right\} \; .
\end{equation}

The appearance of $\sqrt{m_{c}}$ above is only an artifact of the change
of variables; written in terms of $\sigma$, Eq.(21) contains no such term.
The theoretical ratios of the first two Laplace moments (15) and (21) must
now be confronted with a corresponding ratio involving the experimental
data on the $J/\psi$ system. We parametrize the latter by a sum of
two narrow resonances below $D\bar{D}$ threshold, followed by
a hadronic continuum modelled by perturbative QCD, which gives

\begin{equation}
\Pi (\sigma)|_{EXP} = \frac{27}{16 \pi} \frac{1}{\alpha_{EM}^{2}} \sum_{V}
\Gamma_{V}^{ee} M_{V} \exp (- \sigma M_{V}^{2})
+ \frac{1}{\pi} \int_{s_{0}}^{\infty} ds \; \exp (- \sigma s) \; \mbox{Im}
\; \Pi (s)|_{QCD} \; .
\end{equation}

The experimental ratio is then calculated using (22) in (5). The
continuum threshold $s_{0}$ is chosen at or below the $D\bar{D}$
threshold. Reasonable changes in the value of $s_{0}$ have essentially
no impact on the results, as $\Pi(\sigma)$ is saturated almost
entirely by the first two $J/\psi$ narrow resonances.

In the theoretical ratios we use the current values: $\Lambda = 200 - 300
\; \mbox{MeV}$ , for four flavours \cite{ref2}, and
$<\alpha_{s} G^{2}> = 0.063 -
 0.19 \;\mbox{GeV}^{4}$  \cite{ref5}. We find that
theoretical and experimental
ratios match in the wide sum rule window: $\sigma \simeq 0.8 - 1.5 \;$
$\;\mbox{GeV}^{- 2}$, for $m_{c} = 1.39 - 1.46 \;\mbox{GeV}$ in the fully
relativistic case; and  $\sigma \simeq 0.6 - 0.8 \;\mbox{GeV}^{-2}$,
$m_{c} = 1.40 - 1.53 \;\mbox{GeV}$ in the non-relativistic case.
Figure 1 shows the behaviour of (15) for $\Lambda = 200 \;\mbox{MeV}$, and\\
$<\alpha_{s} G^{2}> = 0.063 \;\mbox{GeV}^{4}$ (solid curve),
together with the experimental ratio calculated from (22) (broken curve),
corresponding to $m_{c} = 1.44 \;\mbox{GeV}$. A similar qualitative behaviour
is obtained for other values of $\Lambda$ and the gluon condensate, both
in the relativistic and the non-relativistic versions of the sum rules.
For values of $\sigma$ inside the sum rule window, the hierarchy of the
various terms in the non-relativistic Laplace ratio (21) guarantees a
fast convergence. In fact, the leading correction in $1/m_{c}$ is at
the 15-20\% level, the radiative correction and the non-perturbative
contribution amount both to less than 10\% . At the same time,
the next, and next-to-next to leading (in $1/m_{c}$) terms everywhere in
(21) are safely small, as it can be easily verified from (21) noticing
that if $\sigma \simeq 1/2 \;\mbox{GeV}^{-2}$, then $\tau \simeq 2 m_{c}$.
Clearly, the complete analysis at the level of accuracy of these
next-to-leading mass corrections would require the
evaluation of the perturbative $O(\alpha_s^2)$ terms.
Combining the results from both versions of the Laplace ratios, leads to
the result
\begin{equation}
m_{c} (Q^{2} = m_{c}^{2}) = 1.46 \pm 0.07 \; \mbox{\mbox{\mbox{GeV}}}\; .
\end{equation}
\\
In order to facilitate the comparison of (23) with previous determinations
based on various versions of QCD sum rules \cite{ref3},
\cite{ref8} - \cite{ref14},
we show in Fig. (2) the dependence of $m_{c}$ on
$\Lambda$ for  three different values of $<\alpha_{s} G^{2}>$,
and in Fig. (3) the dependence of $m_{c}$ on the gluon condensate for
$\Lambda$ in the range: $\Lambda = 100 - 400 \;\mbox{MeV}$.
Both figures correspond
to the fully relativistic version of the QCD sum rules. Figures 4 and 5
refer to the non-relativistic determination.
In comparing values of $m_{c}$ from different determinations, it is
important to know which values of $\Lambda$ and $<\alpha_{s} G^{2}>$
have been used, as well as which renormalization point has been chosen,
e.g. some authors determine $m_{c}(Q^{2}=- m_{c}^{2})$, which is related to
the on-shell mass $m_{c}(m_{c}^{2})$ through
\begin{equation}
m_{c}^{2}(m_{c}^{2}) = m_{c}^{2}(-m_{c}^{2})
(1 + \frac{4 \; \ln \; 2}{\pi} \alpha_{s}) \; .
\end{equation}
After using the same values of $\Lambda$ and $<\alpha_{s} G^{2}>$ as used
in \cite{ref3}, \cite{ref8}-\cite{ref14}, and after converting to the
on-shell mass (if necessary), we find that results from the present method
are in very good agreement with those of \cite{ref3} and \cite{ref8}, and
agree within errors with \cite{ref9}-\cite{ref11}; the latter determinations
being on the low side of our error bars.
On the other hand,
the technique used here gives values of $m_{c}$ somewhat
higher than those obtained in  \cite{ref12}-\cite{ref14}.

Recently \cite{ref15},
the two-loop correction to the Wilson coefficient of the gluon condensate
has been calculated. We have incorporated this additional term in the
Laplace ratios, and find that the correction it introduces is about a factor
of 2 larger than the $1/m_{c}$ correction to $<\alpha_{s} G^{2}>$
but with an opposite sign. Given
the relative smallness of the overall contribution from the gluon condensate,
and the conservatively large uncertainty we have allowed in its value, the
final result for $m_{c}$ remains basically unchanged.

Finally, regarding the recent observation that the definition of
heavy quark pole mass in the context of the Heavy Quark Effective Theory (HQET)
should contain some intrinsic ambiguity beyond perturbation theory
\cite{ref16}, due to nonperturbative long distance effects,
we notice that the hadronic
system considered here is made of two heavy quarks, while HQET strictly
applies to heavy-light bound states.\footnote{An application of heavy quark
symmetry to a heavy quark-antiquark system has been presented in \cite{ref17}.}
The non-relativistic ratio (21) is not a relation of the HQET, although it
is obtained formally from the relativistic ratio (15) in the large
quark-mass limit. The purpose of considering (21) together with (15)
has been to assess the size of relativistic corrections, as well as
of systematic uncertainties of QCD sum rules. The numerical results, and
in particular the consistency between the relativistic and nonrelativistic
determinations of $m_c$, indicate that these effects should be small.

{\bf Acknowledgements}\\
The work of (CAD) and (GRG) was  supported in part by the Foundation for
Research Development. The work of (NP) is supported in part by the Human
Capital and Mobility Programme, EEC Contract ERBCHRXCT930132. A discussion
with P. Colangelo is gratefully acknowledged. \\

\subsection*{Figure Captions}
\begin{itemize}
\item[Figure 1:] The fully relativistic ratio (15) (solid curve), and
the experimental ratio (broken curve). The values of the parameters
are: $\Lambda =$ 200 MeV, $\langle\alpha_{s} \; G^{2} \rangle =$ 0.063
GeV$^{4}$, and $m_{c} =$ 1.44 GeV.
\item[Figure 2:] Dependence of $m_{c}$ on $\Lambda$ for $\langle
\alpha_{s} \; G^{2} \rangle =$  0.038 GeV$^{4}$ (curve (a)),
0.063 GeV$^{4}$
(curve (b)), and 0.19 GeV$^{4}$ (curve (c)). The fully relativistic ratio
(15) has been used. Curve (a) is shown for reference purposes, as
this low value of $\langle \alpha_{s} \; G^{2} \rangle $ was not
used in our analysis.
\item[Figure 3:] Dependence of $m_{c}$ on $\langle \alpha_{s} \;
G^{2} \rangle$ for $\Lambda =$ 100 MeV (curve (a)), 200 MeV (curve (b)),
300 MeV (curve (c)), and 400 MeV (curve (d)). The fully relativistic ratio
(15) has been used. Curves (a) and (d) are shown for reference purposes, as
these extreme values of $\Lambda$ were not used in our analysis.
\item[Figure 4:] Same as in Fig.2, except that the non-relativistic
ratio (21) has been used.
\item[Figure 5:] Same as in Fig.3, except that the non-relativistic
ratio (21) has been used.
\end{itemize}
\end{document}